\documentclass[aps,twocolumn,amsfonts,showpacs,superscriptaddress,noshowpacs ]{revtex4}
\usepackage{amssymb,amsmath,amsthm,booktabs,mathtools}
\usepackage{bm}
\usepackage{color}
\usepackage{tikz}
\usepackage{enumitem}
\usepackage{units}
\usepackage{braket}
\usepackage{amsmath}
\usepackage{cancel}

\usepackage{soul} 
\usepackage{units}

\usepackage{braket}

\newcommand{\Gammatot}{\Gamma}  

\begin{document}

\title{Tailoring photon statistics with an atom-based two-photon interferometer}


\author{Martin Cordier}
\thanks{These two authors contributed equally}
\author{Max Schemmer}
\thanks{These two authors contributed equally}
\author{Philipp Schneeweiss}
\author{Jürgen Volz}
\author{Arno Rauschenbeutel}
\email{arno.rauschenbeutel@hu-berlin.de}
\affiliation{Department of Physics, Humboldt-Universit\"{a}t zu Berlin, 10099 Berlin, Germany}
\date{\today}



\maketitle


\textbf{Controlling the photon statistics of light is paramount for quantum science and technologies. 
Recently, we demonstrated that transmitting resonant laser light past an ensemble of two-level emitters can result in a stream of single photons or excess photon pairs. This transformation is due to quantum interference between the transmitted and incoherently scattered two-photon component~\cite{prasad_correlating_2020-1}. Here, using the dispersion of the atomic medium, we actively control the relative quantum phase between these two components. We thereby realize a tunable two-photon interferometer and observe interference fringes in the normalized photon coincidence rate, varying from antibunching to bunching. Beyond the fundamental insight that the quantum phase between incoherent and coherent light can be tuned and dictates photon statistics, our results lend themselves to the development of novel quantum light sources.}

-------------------------------------------------------------------

Non-classical light is a resource in science and technology with applications ranging from quantum communication~\cite{yin_satellite-based_2017} and information processing~\cite{zhong_quantum_2020} to sensing~\cite{mitchell_super-resolving_2004}, imaging~\cite{gatto_monticone_beating_2014,kviatkovsky_microscopy_2020} and metrology~\cite{aasi_enhanced_2013-1}. Depending on the respective application, very different photon statistics are required, ranging from streams of single photons to photon pairs. While the former are temporarily anticorrelated and thus exhibit antibunching, an excess of photon pairs manifests itself as time-correlated or bunched detection events.   
Non-linear media are an important tool for generating such non-classical light and are  used, e.g., in down-conversion based quantum light sources~\cite{wang_multidimensional_2018-1, zhong_12-photon_2018}. The strongest optical non-linearity is granted by quantum emitters like atoms, molecules, color-centers, or quantum dots. Quantum emitters even allow one to reach the regime of quantum non-linear optics, where the response of the medium to the incident light already strongly differs between one and two incident photons. Thanks to this property, nowadays, two-level quantum emitters are widely used to generate antibunched light~\cite{senellart_high-performance_2017-2,lenzini_diamond_2018, rezai_coherence_2018}.

Already about forty years ago, it was conjectured that antibunching in the resonance fluorescence of a two-level emitter stems from fully destructive quantum interference between incoherently and coherently scattered photons~\cite{dalibard_correlation_1983}.
Building on this insight, it has been recently shown that when rejecting one of the components, the photon statistics of the remaining fluorescence light of a single quantum emitter can be modified, from antibunching to Poissonian distribution~\cite{hanschke_origin_2020-4,phillips_photon_2020} or to bunching~\cite{masters_will_2022-1,ng_observation_2022-1}. 
Moreover, it has been shown that a similar quantum interference effect occurs when transmitting resonant light past an ensemble of two-level emitters~\cite{mahmoodian_strongly_2018,prasad_correlating_2020-1}. In all these demonstrations, only the relative amplitude between the coherent and incoherent light was modified, thus only reducing the contrast of their destructive interference. 

Here, we take an important step forward on these previous works and demonstrate full control over the relative phase and amplitude of the coherent and incoherent light that is transmitted through an ensemble of laser-cooled atoms.
Taking advantage of the dispersion of this atomic medium, we realize a two-photon interferometer that allows us to continuously tune the nature of the interference from destructive to constructive. 
Scanning the phase of this interferometer, we observe interference fringes in the photon coincidence rate, thereby supporting the fundamental insight that the quantum phase between the coherent and incoherent light dictates the photon statistics. 

\begin{figure}[t!]
    \centering
    \includegraphics[width=1\linewidth]{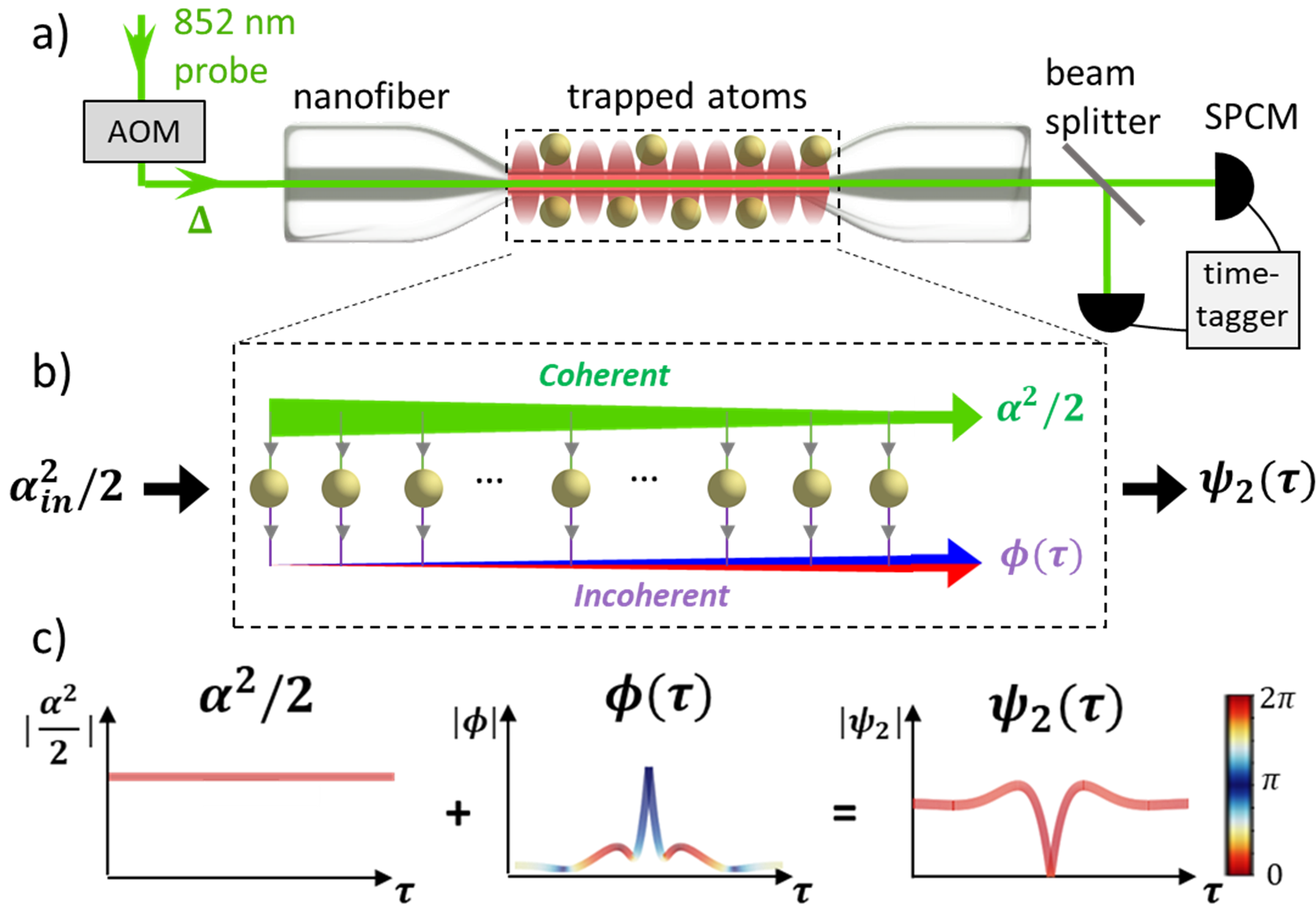}
    \caption{a) Experimental setup. Laser-cooled Cs atoms, which are optically trapped in the evanescent field surrounding an optical nanofiber, are probed with a near-resonant CW laser (green). The detuning between the probe and the atomic resonance, $\Delta$, is set by an acousto-optic modulator (AOM). After the interaction with the atomic ensemble, the transmitted light contains a coherent and incoherent component. The photon statistic of the transmitted light is measured with a Hanbury-Brown and Twiss setup. b) Sketch of the effective multipath interferometer for the coherent (upper arm) and incoherent (lower arm) two-photon component. While the two components are sketched separately, they propagate inside the same spatial mode of the nanofiber.  c) Example of the two-photon wavefunction $\psi_2(\tau)$ resulting from an interference between the coherent ($\alpha^2/2$) and the incoherent ($\phi$) two-photon complex amplitudes.} 
    \label{fig_interferometer}
\end{figure}

Our setup is sketched in Fig.~\ref{fig_interferometer} a). It comprises an optical nanofiber through which a continuous wave (CW) probe laser field is launched with photon flux $|\alpha_\mathrm{in}|^2$ and frequency $\omega_\mathrm{L}$. In the nanofiber region, the probe light protrudes beyond the nanofiber surface in the form of an evanescent field which couples to an ensemble of $N$ laser-cooled cesium atoms trapped along the nanofiber (D2-line: resonance frequency $\omega_0$, natural linewidth $\Gamma$). The photon statistics of the transmitted light is measured with a Hanbury-Brown and Twiss setup (HBT). Each of the atoms scatters the incident light in two different ways, which are referred to as coherent and incoherent scattering. 
This denomination reflects their respective capability to interfere with the electric field of the excitation laser. 
In the low saturation regime, incoherent scattering is a process that resembles spontaneous four-wave mixing where a pair of frequency entangled red- and blue-detuned photons are created~\cite{dalibard_correlation_1983,le_jeannic_experimental_2021,hinney_unraveling_2021-2,shen_strongly_2007-1}. In forward direction, the incoherently scattered components from different atoms add up coherently which gives rise to a collective enhancement~\cite{mahmoodian_strongly_2018,prasad_correlating_2020-1}.
At the same time, the coherently forward scattered components interfere destructively with the probe light leading to an exponential attenuation of the latter with the number of atoms,  according to Beer-Lambert's law. 
For small atom numbers, the photon statistics is dominated by the transmitted probe light, thus resembling that of a coherent state. For larger atom numbers, however, the combined effect of exponential attenuation of the probe and the collective build-up of the incoherent component gives rise to modified photon statistics. Depending on $N$ and laser-atom detuning, $\Delta = \omega_L - \omega_0$, the interference of the two components yields bunching or antibunching in the second-order correlation function, $g^{(2)}(\tau)$.

\begin{figure}[t]
\centering
\includegraphics[width=\linewidth]{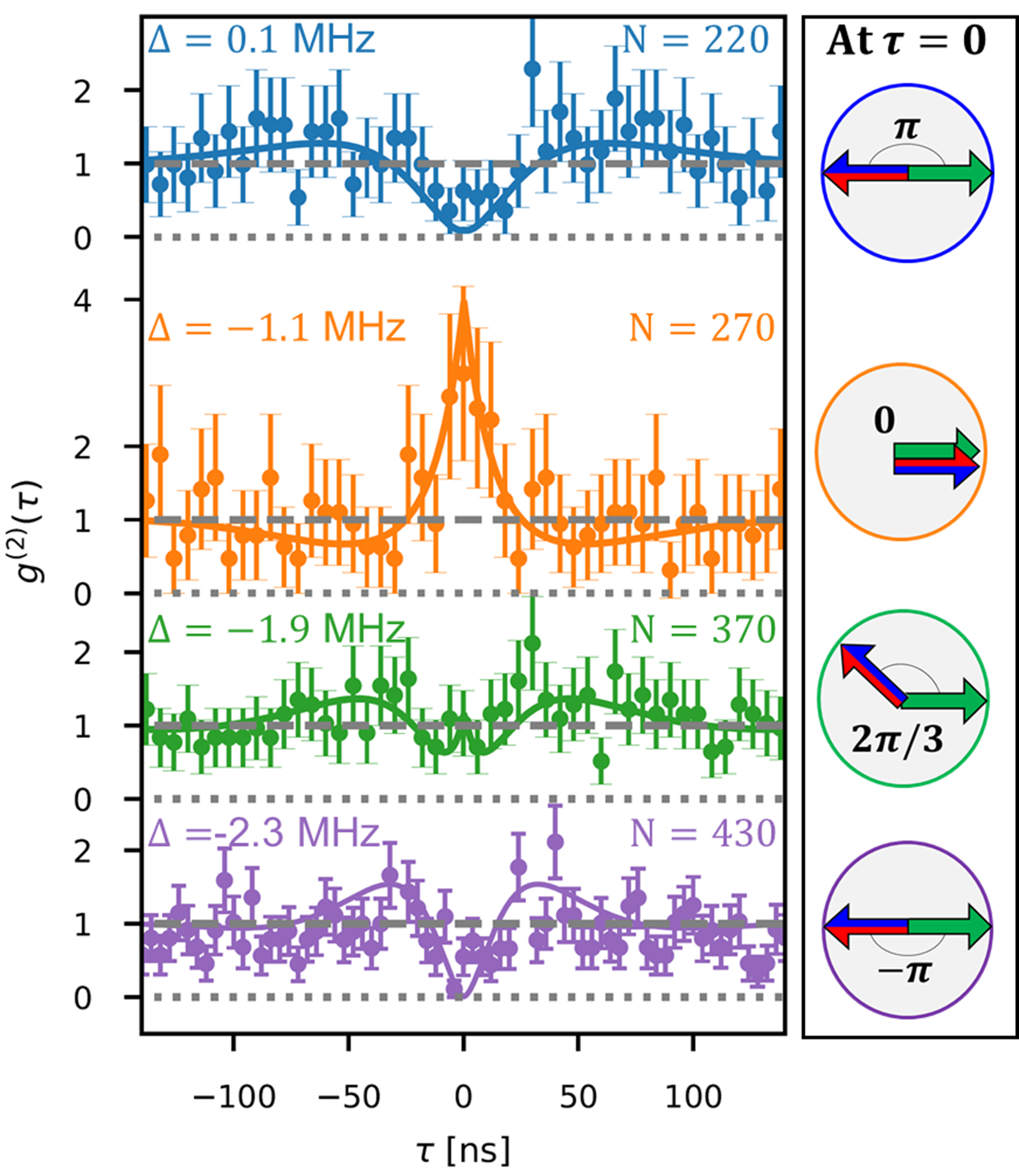}
\caption{Measured second-order correlation $g^{(2)}(\tau)$ for different relative phases $\Delta\varphi(0)$. In each configuration, the number of atoms is adjusted to reach an equal amplitude of coherent and incoherent two-photon component at zero delay. The solid lines correspond to the model prediction. 
The two arrows in the right panel represent the complex values of the coherent (green) and the incoherent (blue/red) two-photon amplitude at $\tau =0$ in each configuration.}
\label{fig_g2_OD}
\end{figure}

To quantitatively describe this effect, one has to consider the quantum state of the two-photon component of the transmitted light~\cite{mahmoodian_strongly_2018}: 
\begin{equation}
    \ket{\psi_2} = \iint dt_1 dt_2 \, \psi_2(t_2-t_1) \hat{a}^\dagger_{t_1} \hat{a}^\dagger_{t_2} \ket{0}
\end{equation}
where $\hat{a}^\dagger_{t}$ creates a photon at time $t$. The temporal wavefunction, $\psi_2(\tau)$, gives the probability amplitude to detect two photons with a time delay $\tau = t_2 - t_1$. In the low saturation regime, this two-photon wavefunction can be expressed as
\begin{equation}
\psi_2(\tau) = \alpha^2/2+  \phi(\tau)
\end{equation}
where $\alpha^2/2$ is the two-photon wavefunction of the attenuated probe light and $\phi(\tau)$ is the wavefunction of collectively scattered incoherent component. \nopagebreak{Here, $\alpha= t^{N}(\Delta) \alpha_\mathrm{in}$ }where $t(\Delta)$ is the complex amplitude transmission coefficient of a single atom, see SM. In the case of CW excitation, $\alpha^2/2$ is constant, thereby describing uncorrelated pairs of photons. In contrast, the incoherent two-photon amplitude $\phi$ depends on $\tau$ due to temporal correlations induced by local atom-mediated photon-photon interactions. These correlations thus extend over the excited state lifetime, $1/\Gamma$~\cite{shen_strongly_2007-1,mahmoodian_strongly_2018}. We calculate $\phi(\tau)$ using a model that builds on the work in Ref.~\cite{mahmoodian_strongly_2018}, see SM. The resulting photon statistics of the transmitted light is then proportional to the modulus square of the two-photon wavefunction, $g^{(2)}(\tau) \propto |\psi_2(\tau)|^2$. Thus, $g^{(2)}(\tau)$ can be interpreted as an interference signal of a two-photon interferometer as sketched in Fig.~\ref{fig_interferometer} b). In this picture, each atom acts as a nonlinear beamsplitter that coherently splits the incoming light into a coherently and an incoherently scattered component. At the same time, each atom also imparts a relative phase shift between the coherent and incoherent component. As a consequence, after the interaction with the ensemble, the interfering complex amplitudes, $\alpha^2/2$ and $\phi(\tau)$, depend both in magnitude and phase on the detuning of the probe light and the number of atoms. In general, the phase difference between the two components also depend on $\tau$.  
Experimentally, the relative phase $\Delta\varphi(\tau) = \arg\lbrace \phi(\tau)\rbrace - \arg\lbrace \alpha^2/2\rbrace$ and amplitude $|\phi(\tau)|/|\alpha^2/2|$ of the two components are controlled by adjusting the detuning $\Delta$ and the atom number $N$, see Fig.~\ref{fig_interferometer} a) and SM. 

\begin{figure}[t]
\centering
\includegraphics[width=\linewidth]{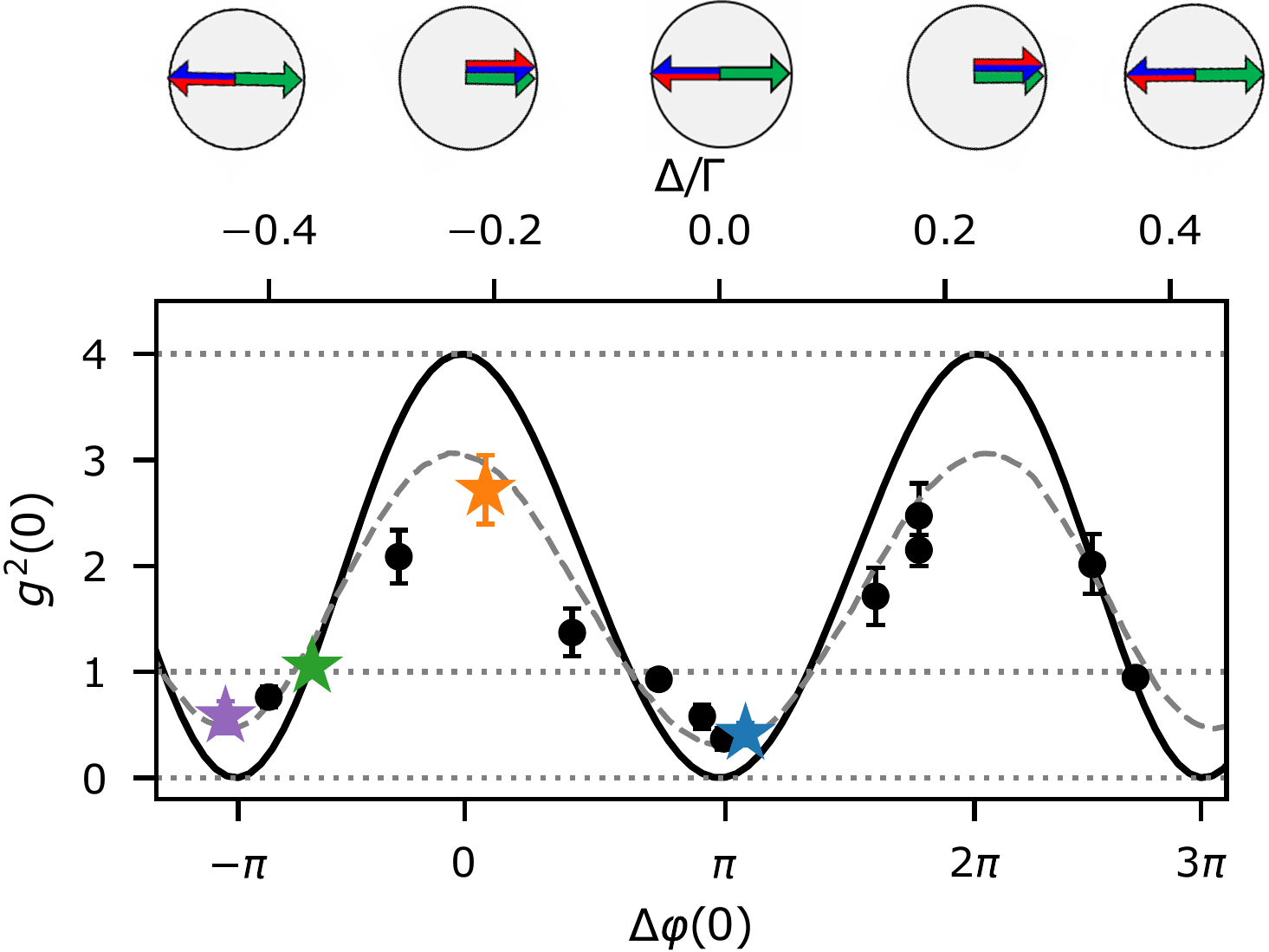}
\caption{Interference fringes in the normalized second-order correlation $g^{(2)}(\tau = 0)$ as a function of the phase difference between coherent and incoherent photon pairs $\Delta \varphi (0)$. All points are taken for equal coherent and incoherent amplitude at zero delay. The detuning $\Delta$ is shown in the top x-axis. The solid line is the model prediction assuming ideal conditions, while the dashed line takes into the effect of averaging over a finite range of $N$ (see SM). 
The four data points indicated as stars correspond to the data shown in Fig.~\ref{fig_g2_OD}.}
\label{fig_am}
\end{figure}

Figure~\ref{fig_g2_OD} shows measured second-order correlation functions, $g^{(2)}(\tau)$,  for four different values of $\Delta$. 
For each detuning $\Delta$, the atom number was chosen such as to yield equal magnitudes of the coherent and incoherent two-photon amplitude at zero time delay, i.e., $|\alpha^2/2| = |\phi(\tau = 0)|$, thereby maximizing their interference contrast. 
The right column in Fig.~\ref{fig_g2_OD} shows $\alpha^2/2$ and $\phi(\tau=0)$ in the complex plane with green and red/blue arrows, respectively. Here, $\alpha^2/2$ is taken as the phase reference and is thus assumed to be real and positive. On resonance (first row), their relative phase is $\Delta\varphi(0) = \pi$ leading to destructive interference between the two wavefunctions. This gives rise to photon antibunching in the measured correlation function reaching $g^{(2)}(0)= 0.4 \pm 0.1$ (blue data).
When the laser light is detuned, the phase shift $\Delta\varphi(0)$ increases gradually. As a consequence, for $\Delta = -1.1 \,\mathrm{MHz}$ (second row), the wavefunctions are in phase and constructively interfere at $\tau = 0$. This results in photon bunching of $g^{(2)}(0)= 2.7 \pm 0.3$, somewhat lower than $g^{(2)}(0)=4$, expected for fully constructive interference and equal amplitudes.
In the third row, the detuning $\Delta = -1.9 \,\mathrm{MHz}$ was chosen to yield a phase shift $\Delta\varphi(0) = 2\pi/3$ and, accordingly, $g^{(2)}(0)=1.05\pm 0.02$. At a detuning $\Delta= -2.3 \,\mathrm{MHz}$, as shown in the fourth row,  the two components are again $\pi$ out of phase. The resulting resurrection of antibunching for finite detuning is remarkable and clearly illustrates the interference character of the observed phenomenon. 
We note that the oscillations of $g^{(2)}$ around 1 for finite $\tau$ stem from the $\tau$-dependent relative phase between $\alpha^2/2$ and $\phi(\tau)$, see center panel in Fig.~\ref{fig_interferometer}.c. 

We now repeat the above measurement procedure for 15 different relative phases $\Delta\varphi(0)$ and use a maximum likelihood estimation to extract the normalized photon coincidences $g^{(2)}(0)$ (SM). The results of this analysis are plotted in Fig.~\ref{fig_am}. When modifying the phase difference, the photon statistics oscillates between antibunching and bunching. For comparison, the solid black line shows the theoretical prediction for ideal conditions, which oscillates sinusoidally between perfect antibunching for $\Delta\varphi(0)  = \pm \pi n$ ($n= 1,3,\ldots$) and photon bunching with  $g^{(2)}(0) = 4$ for $\Delta\varphi(0)  = \pm \pi n$ ($n= 0,2,4,\ldots$). In comparison to this ideal case, our experimental data show a slightly reduced visibility $V = 0.76 \pm 0.02$ whose value is determined from the measured maximum and minimum $g^{(2)}(0)$. The dashed gray line  is a theoretical prediction including the effect of averaging over a finite range in $N$ (see supplemental).


The observed interference fringes obtained by tuning the relative phase and amplitude of the coherent and incoherent component demonstrate the coherence between the coherent and incoherent two-photon amplitudes, both in the resonant and detuned case. 
 
In conclusion, our results show that the second-order quantum correlation function, $g^{(2)}(\tau)$, of near-resonant light transmitted past an ensemble of $N$ two-level emitters can be quantitatively understood as resulting from the interference of the coherent and incoherent two-photon amplitude in an atom-based two-photon interferometer. This insight and the demonstrated control over the relative amplitude and phase open a new avenue toward tailoring the photon statistics of laser light for application in the realm of quantum technologies. In that context, while the current study focuses on the interference of the two-photon component, the same interference mechanism is expected to occur for larger photon-number components. 
This may then pave the way for the generation and engineering of quantum states of light containing large photon-numbers.\\  %




\textbf{Acknowledgements}
We are grateful to J. Dalibard for his feedback and careful reading of the manuscript. 
M. C. and M. S acknowledge support from the European Commission (Marie Skłodowska-Curie IF Grant No.~101029304 and IF Grant No.~896957). We acknowledge funding by the Alexander von Humboldt Foundation in the framework of the Alexander von Humboldt Professorship endowed by the Federal Ministry of Education and Research, as well as funding by the European Commission under the project DAALI (No.~899275).

\bibliography{higher_order2}

\begin{thebibliography}{30}
\expandafter\ifx\csname natexlab\endcsname\relax\def\natexlab#1{#1}\fi
\expandafter\ifx\csname bibnamefont\endcsname\relax
  \def\bibnamefont#1{#1}\fi
\expandafter\ifx\csname bibfnamefont\endcsname\relax
  \def\bibfnamefont#1{#1}\fi
\expandafter\ifx\csname citenamefont\endcsname\relax
  \def\citenamefont#1{#1}\fi
\expandafter\ifx\csname url\endcsname\relax
  \def\url#1{\texttt{#1}}\fi
\expandafter\ifx\csname urlprefix\endcsname\relax\def\urlprefix{URL }\fi
\providecommand{\bibinfo}[2]{#2}
\providecommand{\eprint}[2][]{\url{#2}}

\bibitem[{\citenamefont{Prasad et~al.}(2020)\citenamefont{Prasad, Hinney,
  Mahmoodian, Hammerer, Rind, Schneeweiss, S{\o}rensen, Volz, and
  Rauschenbeutel}}]{prasad_correlating_2020-1}
\bibinfo{author}{\bibfnamefont{A.~S.} \bibnamefont{Prasad}},
  \bibinfo{author}{\bibfnamefont{J.}~\bibnamefont{Hinney}},
  \bibinfo{author}{\bibfnamefont{S.}~\bibnamefont{Mahmoodian}},
  \bibinfo{author}{\bibfnamefont{K.}~\bibnamefont{Hammerer}},
  \bibinfo{author}{\bibfnamefont{S.}~\bibnamefont{Rind}},
  \bibinfo{author}{\bibfnamefont{P.}~\bibnamefont{Schneeweiss}},
  \bibinfo{author}{\bibfnamefont{A.~S.} \bibnamefont{S{\o}rensen}},
  \bibinfo{author}{\bibfnamefont{J.}~\bibnamefont{Volz}}, \bibnamefont{and}
  \bibinfo{author}{\bibfnamefont{A.}~\bibnamefont{Rauschenbeutel}},
  \bibinfo{journal}{Nature Photonics}  (\bibinfo{year}{2020}).

\bibitem[{\citenamefont{Yin et~al.}(2017)\citenamefont{Yin, Cao, Li, Liao,
  Zhang, Ren, Cai, Liu, Li, Dai et~al.}}]{yin_satellite-based_2017}
\bibinfo{author}{\bibfnamefont{J.}~\bibnamefont{Yin}},
  \bibinfo{author}{\bibfnamefont{Y.}~\bibnamefont{Cao}},
  \bibinfo{author}{\bibfnamefont{Y.-H.} \bibnamefont{Li}},
  \bibinfo{author}{\bibfnamefont{S.-K.} \bibnamefont{Liao}},
  \bibinfo{author}{\bibfnamefont{L.}~\bibnamefont{Zhang}},
  \bibinfo{author}{\bibfnamefont{J.-G.} \bibnamefont{Ren}},
  \bibinfo{author}{\bibfnamefont{W.-Q.} \bibnamefont{Cai}},
  \bibinfo{author}{\bibfnamefont{W.-Y.} \bibnamefont{Liu}},
  \bibinfo{author}{\bibfnamefont{B.}~\bibnamefont{Li}},
  \bibinfo{author}{\bibfnamefont{H.}~\bibnamefont{Dai}}, \bibnamefont{et~al.},
  \bibinfo{journal}{Science} \textbf{\bibinfo{volume}{356}},
  \bibinfo{pages}{1140} (\bibinfo{year}{2017}).

\bibitem[{\citenamefont{Zhong et~al.}(2020)\citenamefont{Zhong, Wang, Deng,
  Chen, Peng, Luo, Qin, Wu, Ding, Hu et~al.}}]{zhong_quantum_2020}
\bibinfo{author}{\bibfnamefont{H.-S.} \bibnamefont{Zhong}},
  \bibinfo{author}{\bibfnamefont{H.}~\bibnamefont{Wang}},
  \bibinfo{author}{\bibfnamefont{Y.-H.} \bibnamefont{Deng}},
  \bibinfo{author}{\bibfnamefont{M.-C.} \bibnamefont{Chen}},
  \bibinfo{author}{\bibfnamefont{L.-C.} \bibnamefont{Peng}},
  \bibinfo{author}{\bibfnamefont{Y.-H.} \bibnamefont{Luo}},
  \bibinfo{author}{\bibfnamefont{J.}~\bibnamefont{Qin}},
  \bibinfo{author}{\bibfnamefont{D.}~\bibnamefont{Wu}},
  \bibinfo{author}{\bibfnamefont{X.}~\bibnamefont{Ding}},
  \bibinfo{author}{\bibfnamefont{Y.}~\bibnamefont{Hu}}, \bibnamefont{et~al.},
  \bibinfo{journal}{Science} \textbf{\bibinfo{volume}{370}},
  \bibinfo{pages}{1460} (\bibinfo{year}{2020}).

\bibitem[{\citenamefont{Mitchell et~al.}(2004)\citenamefont{Mitchell, Lundeen,
  and Steinberg}}]{mitchell_super-resolving_2004}
\bibinfo{author}{\bibfnamefont{M.~W.} \bibnamefont{Mitchell}},
  \bibinfo{author}{\bibfnamefont{J.~S.} \bibnamefont{Lundeen}},
  \bibnamefont{and} \bibinfo{author}{\bibfnamefont{A.~M.}
  \bibnamefont{Steinberg}}, \bibinfo{journal}{Nature}
  \textbf{\bibinfo{volume}{429}}, \bibinfo{pages}{161} (\bibinfo{year}{2004}).

\bibitem[{\citenamefont{Gatto~Monticone
  et~al.}(2014)\citenamefont{Gatto~Monticone, Katamadze, Traina, Moreva,
  Forneris, {Ruo-Berchera}, Olivero, Degiovanni, Brida, and
  Genovese}}]{gatto_monticone_beating_2014}
\bibinfo{author}{\bibfnamefont{D.}~\bibnamefont{Gatto~Monticone}},
  \bibinfo{author}{\bibfnamefont{K.}~\bibnamefont{Katamadze}},
  \bibinfo{author}{\bibfnamefont{P.}~\bibnamefont{Traina}},
  \bibinfo{author}{\bibfnamefont{E.}~\bibnamefont{Moreva}},
  \bibinfo{author}{\bibfnamefont{J.}~\bibnamefont{Forneris}},
  \bibinfo{author}{\bibfnamefont{I.}~\bibnamefont{{Ruo-Berchera}}},
  \bibinfo{author}{\bibfnamefont{P.}~\bibnamefont{Olivero}},
  \bibinfo{author}{\bibfnamefont{I.~P.} \bibnamefont{Degiovanni}},
  \bibinfo{author}{\bibfnamefont{G.}~\bibnamefont{Brida}}, \bibnamefont{and}
  \bibinfo{author}{\bibfnamefont{M.}~\bibnamefont{Genovese}},
  \bibinfo{journal}{Physical Review Letters} \textbf{\bibinfo{volume}{113}},
  \bibinfo{pages}{143602} (\bibinfo{year}{2014}).

\bibitem[{\citenamefont{Kviatkovsky et~al.}(2020)\citenamefont{Kviatkovsky,
  Chrzanowski, Avery, Bartolomaeus, and Ramelow}}]{kviatkovsky_microscopy_2020}
\bibinfo{author}{\bibfnamefont{I.}~\bibnamefont{Kviatkovsky}},
  \bibinfo{author}{\bibfnamefont{H.~M.} \bibnamefont{Chrzanowski}},
  \bibinfo{author}{\bibfnamefont{E.~G.} \bibnamefont{Avery}},
  \bibinfo{author}{\bibfnamefont{H.}~\bibnamefont{Bartolomaeus}},
  \bibnamefont{and} \bibinfo{author}{\bibfnamefont{S.}~\bibnamefont{Ramelow}},
  \bibinfo{journal}{Science Advances} \textbf{\bibinfo{volume}{6}}
  (\bibinfo{year}{2020}).

\bibitem[{\citenamefont{Aasi et~al.}(2013)\citenamefont{Aasi, Abadie, Abbott,
  Abbott, Abbott, Abernathy, Adams, Adams, Addesso, Adhikari
  et~al.}}]{aasi_enhanced_2013-1}
\bibinfo{author}{\bibfnamefont{J.}~\bibnamefont{Aasi}},
  \bibinfo{author}{\bibfnamefont{J.}~\bibnamefont{Abadie}},
  \bibinfo{author}{\bibfnamefont{B.~P.} \bibnamefont{Abbott}},
  \bibinfo{author}{\bibfnamefont{R.}~\bibnamefont{Abbott}},
  \bibinfo{author}{\bibfnamefont{T.~D.} \bibnamefont{Abbott}},
  \bibinfo{author}{\bibfnamefont{M.~R.} \bibnamefont{Abernathy}},
  \bibinfo{author}{\bibfnamefont{C.}~\bibnamefont{Adams}},
  \bibinfo{author}{\bibfnamefont{T.}~\bibnamefont{Adams}},
  \bibinfo{author}{\bibfnamefont{P.}~\bibnamefont{Addesso}},
  \bibinfo{author}{\bibfnamefont{R.~X.} \bibnamefont{Adhikari}},
  \bibnamefont{et~al.}, \bibinfo{journal}{Nature Photonics}
  \textbf{\bibinfo{volume}{7}}, \bibinfo{pages}{613} (\bibinfo{year}{2013}).

\bibitem[{\citenamefont{Wang et~al.}(2018)\citenamefont{Wang, Paesani, Ding,
  Santagati, Skrzypczyk, Salavrakos, Tura, Augusiak, Man{\v c}inska, Bacco
  et~al.}}]{wang_multidimensional_2018-1}
\bibinfo{author}{\bibfnamefont{J.}~\bibnamefont{Wang}},
  \bibinfo{author}{\bibfnamefont{S.}~\bibnamefont{Paesani}},
  \bibinfo{author}{\bibfnamefont{Y.}~\bibnamefont{Ding}},
  \bibinfo{author}{\bibfnamefont{R.}~\bibnamefont{Santagati}},
  \bibinfo{author}{\bibfnamefont{P.}~\bibnamefont{Skrzypczyk}},
  \bibinfo{author}{\bibfnamefont{A.}~\bibnamefont{Salavrakos}},
  \bibinfo{author}{\bibfnamefont{J.}~\bibnamefont{Tura}},
  \bibinfo{author}{\bibfnamefont{R.}~\bibnamefont{Augusiak}},
  \bibinfo{author}{\bibfnamefont{L.}~\bibnamefont{Man{\v c}inska}},
  \bibinfo{author}{\bibfnamefont{D.}~\bibnamefont{Bacco}},
  \bibnamefont{et~al.}, \bibinfo{journal}{Science}
  \textbf{\bibinfo{volume}{360}}, \bibinfo{pages}{285} (\bibinfo{year}{2018}).

\bibitem[{\citenamefont{Zhong et~al.}(2018)\citenamefont{Zhong, Li, Li, Peng,
  Su, Hu, He, Ding, Zhang, Li et~al.}}]{zhong_12-photon_2018}
\bibinfo{author}{\bibfnamefont{H.-S.} \bibnamefont{Zhong}},
  \bibinfo{author}{\bibfnamefont{Y.}~\bibnamefont{Li}},
  \bibinfo{author}{\bibfnamefont{W.}~\bibnamefont{Li}},
  \bibinfo{author}{\bibfnamefont{L.-C.} \bibnamefont{Peng}},
  \bibinfo{author}{\bibfnamefont{Z.-E.} \bibnamefont{Su}},
  \bibinfo{author}{\bibfnamefont{Y.}~\bibnamefont{Hu}},
  \bibinfo{author}{\bibfnamefont{Y.-M.} \bibnamefont{He}},
  \bibinfo{author}{\bibfnamefont{X.}~\bibnamefont{Ding}},
  \bibinfo{author}{\bibfnamefont{W.}~\bibnamefont{Zhang}},
  \bibinfo{author}{\bibfnamefont{H.}~\bibnamefont{Li}}, \bibnamefont{et~al.},
  \bibinfo{journal}{Physical Review Letters} \textbf{\bibinfo{volume}{121}},
  \bibinfo{pages}{250505} (\bibinfo{year}{2018}).

\bibitem[{\citenamefont{Senellart et~al.}(2017)\citenamefont{Senellart,
  Solomon, and White}}]{senellart_high-performance_2017-2}
\bibinfo{author}{\bibfnamefont{P.}~\bibnamefont{Senellart}},
  \bibinfo{author}{\bibfnamefont{G.}~\bibnamefont{Solomon}}, \bibnamefont{and}
  \bibinfo{author}{\bibfnamefont{A.}~\bibnamefont{White}},
  \bibinfo{journal}{Nature Nanotechnology} \textbf{\bibinfo{volume}{12}},
  \bibinfo{pages}{1026} (\bibinfo{year}{2017}).

\bibitem[{\citenamefont{Lenzini et~al.}(2018)\citenamefont{Lenzini, Gruhler,
  Walter, and Pernice}}]{lenzini_diamond_2018}
\bibinfo{author}{\bibfnamefont{F.}~\bibnamefont{Lenzini}},
  \bibinfo{author}{\bibfnamefont{N.}~\bibnamefont{Gruhler}},
  \bibinfo{author}{\bibfnamefont{N.}~\bibnamefont{Walter}}, \bibnamefont{and}
  \bibinfo{author}{\bibfnamefont{W.~H.~P.} \bibnamefont{Pernice}},
  \bibinfo{journal}{Advanced Quantum Technologies}
  \textbf{\bibinfo{volume}{1}}, \bibinfo{pages}{1800061}
  (\bibinfo{year}{2018}).

\bibitem[{\citenamefont{Rezai et~al.}(2018)\citenamefont{Rezai, Wrachtrup, and
  Gerhardt}}]{rezai_coherence_2018}
\bibinfo{author}{\bibfnamefont{M.}~\bibnamefont{Rezai}},
  \bibinfo{author}{\bibfnamefont{J.}~\bibnamefont{Wrachtrup}},
  \bibnamefont{and} \bibinfo{author}{\bibfnamefont{I.}~\bibnamefont{Gerhardt}},
  \bibinfo{journal}{Physical Review X} \textbf{\bibinfo{volume}{8}},
  \bibinfo{pages}{031026} (\bibinfo{year}{2018}).

\bibitem[{\citenamefont{Dalibard and
  Reynaud}(1983)}]{dalibard_correlation_1983}
\bibinfo{author}{\bibfnamefont{J.}~\bibnamefont{Dalibard}} \bibnamefont{and}
  \bibinfo{author}{\bibfnamefont{S.}~\bibnamefont{Reynaud}},
  \bibinfo{journal}{Journal de Physique} \textbf{\bibinfo{volume}{44}},
  \bibinfo{pages}{1337} (\bibinfo{year}{1983}).

\bibitem[{\citenamefont{Hanschke et~al.}(2020)\citenamefont{Hanschke,
  Schweickert, Carre{\~n}o, Sch{\"o}ll, Zeuner, Lettner, Casalengua, Reindl,
  {da Silva}, Trotta et~al.}}]{hanschke_origin_2020-4}
\bibinfo{author}{\bibfnamefont{L.}~\bibnamefont{Hanschke}},
  \bibinfo{author}{\bibfnamefont{L.}~\bibnamefont{Schweickert}},
  \bibinfo{author}{\bibfnamefont{J.~C.~L.} \bibnamefont{Carre{\~n}o}},
  \bibinfo{author}{\bibfnamefont{E.}~\bibnamefont{Sch{\"o}ll}},
  \bibinfo{author}{\bibfnamefont{K.~D.} \bibnamefont{Zeuner}},
  \bibinfo{author}{\bibfnamefont{T.}~\bibnamefont{Lettner}},
  \bibinfo{author}{\bibfnamefont{E.~Z.} \bibnamefont{Casalengua}},
  \bibinfo{author}{\bibfnamefont{M.}~\bibnamefont{Reindl}},
  \bibinfo{author}{\bibfnamefont{S.~F.~C.} \bibnamefont{{da Silva}}},
  \bibinfo{author}{\bibfnamefont{R.}~\bibnamefont{Trotta}},
  \bibnamefont{et~al.}, \bibinfo{journal}{Physical Review Letters}
  \textbf{\bibinfo{volume}{125}}, \bibinfo{pages}{170402}
  (\bibinfo{year}{2020}).

\bibitem[{\citenamefont{Phillips et~al.}(2020)\citenamefont{Phillips, Brash,
  McCutcheon, {Iles-Smith}, Clarke, Royall, Skolnick, Fox, and
  Nazir}}]{phillips_photon_2020}
\bibinfo{author}{\bibfnamefont{C.~L.} \bibnamefont{Phillips}},
  \bibinfo{author}{\bibfnamefont{A.~J.} \bibnamefont{Brash}},
  \bibinfo{author}{\bibfnamefont{D.~P.~S.} \bibnamefont{McCutcheon}},
  \bibinfo{author}{\bibfnamefont{J.}~\bibnamefont{{Iles-Smith}}},
  \bibinfo{author}{\bibfnamefont{E.}~\bibnamefont{Clarke}},
  \bibinfo{author}{\bibfnamefont{B.}~\bibnamefont{Royall}},
  \bibinfo{author}{\bibfnamefont{M.~S.} \bibnamefont{Skolnick}},
  \bibinfo{author}{\bibfnamefont{A.~M.} \bibnamefont{Fox}}, \bibnamefont{and}
  \bibinfo{author}{\bibfnamefont{A.}~\bibnamefont{Nazir}},
  \bibinfo{journal}{Physical Review Letters} \textbf{\bibinfo{volume}{125}},
  \bibinfo{pages}{043603} (\bibinfo{year}{2020}).

\bibitem[{\citenamefont{Masters et~al.}(2022)\citenamefont{Masters, Hu,
  Cordier, Maron, Pache, Rauschenbeutel, Schemmer, and
  Volz}}]{masters_will_2022-1}
\bibinfo{author}{\bibfnamefont{L.}~\bibnamefont{Masters}},
  \bibinfo{author}{\bibfnamefont{X.}~\bibnamefont{Hu}},
  \bibinfo{author}{\bibfnamefont{M.}~\bibnamefont{Cordier}},
  \bibinfo{author}{\bibfnamefont{G.}~\bibnamefont{Maron}},
  \bibinfo{author}{\bibfnamefont{L.}~\bibnamefont{Pache}},
  \bibinfo{author}{\bibfnamefont{A.}~\bibnamefont{Rauschenbeutel}},
  \bibinfo{author}{\bibfnamefont{M.}~\bibnamefont{Schemmer}}, \bibnamefont{and}
  \bibinfo{author}{\bibfnamefont{J.}~\bibnamefont{Volz}},
  \emph{\bibinfo{title}{Will a single two-level atom simultaneously scatter two
  photons?}} (\bibinfo{year}{2022}), \eprint{arXiv:2209.02547}.

\bibitem[{\citenamefont{Ng et~al.}(2022)\citenamefont{Ng, Chow, and
  Kurtsiefer}}]{ng_observation_2022-1}
\bibinfo{author}{\bibfnamefont{B.~L.} \bibnamefont{Ng}},
  \bibinfo{author}{\bibfnamefont{C.~H.} \bibnamefont{Chow}}, \bibnamefont{and}
  \bibinfo{author}{\bibfnamefont{C.}~\bibnamefont{Kurtsiefer}},
  \emph{\bibinfo{title}{Observation of the {{Mollow Triplet}} from an optically
  confined single atom}} (\bibinfo{year}{2022}), \eprint{arXiv:2208.06575}.

\bibitem[{\citenamefont{Mahmoodian et~al.}(2018)\citenamefont{Mahmoodian, {\v
  C}epulkovskis, Das, Lodahl, Hammerer, and
  S{\o}rensen}}]{mahmoodian_strongly_2018}
\bibinfo{author}{\bibfnamefont{S.}~\bibnamefont{Mahmoodian}},
  \bibinfo{author}{\bibfnamefont{M.}~\bibnamefont{{\v C}epulkovskis}},
  \bibinfo{author}{\bibfnamefont{S.}~\bibnamefont{Das}},
  \bibinfo{author}{\bibfnamefont{P.}~\bibnamefont{Lodahl}},
  \bibinfo{author}{\bibfnamefont{K.}~\bibnamefont{Hammerer}}, \bibnamefont{and}
  \bibinfo{author}{\bibfnamefont{A.~S.} \bibnamefont{S{\o}rensen}},
  \bibinfo{journal}{Physical Review Letters} \textbf{\bibinfo{volume}{121}},
  \bibinfo{pages}{143601} (\bibinfo{year}{2018}).

\bibitem[{\citenamefont{Le~Jeannic et~al.}(2021)\citenamefont{Le~Jeannic,
  Ramos, Simonsen, Pregnolato, Liu, Schott, Wieck, Ludwig, Rotenberg,
  {Garc{\'i}a-Ripoll} et~al.}}]{le_jeannic_experimental_2021}
\bibinfo{author}{\bibfnamefont{H.}~\bibnamefont{Le~Jeannic}},
  \bibinfo{author}{\bibfnamefont{T.}~\bibnamefont{Ramos}},
  \bibinfo{author}{\bibfnamefont{S.~F.} \bibnamefont{Simonsen}},
  \bibinfo{author}{\bibfnamefont{T.}~\bibnamefont{Pregnolato}},
  \bibinfo{author}{\bibfnamefont{Z.}~\bibnamefont{Liu}},
  \bibinfo{author}{\bibfnamefont{R.}~\bibnamefont{Schott}},
  \bibinfo{author}{\bibfnamefont{A.~D.} \bibnamefont{Wieck}},
  \bibinfo{author}{\bibfnamefont{A.}~\bibnamefont{Ludwig}},
  \bibinfo{author}{\bibfnamefont{N.}~\bibnamefont{Rotenberg}},
  \bibinfo{author}{\bibfnamefont{J.~J.} \bibnamefont{{Garc{\'i}a-Ripoll}}},
  \bibnamefont{et~al.}, \bibinfo{journal}{Physical Review Letters}
  \textbf{\bibinfo{volume}{126}}, \bibinfo{pages}{023603}
  (\bibinfo{year}{2021}).

\bibitem[{\citenamefont{Hinney et~al.}(2021)\citenamefont{Hinney, Prasad,
  Mahmoodian, Hammerer, Rauschenbeutel, Schneeweiss, Volz, and
  Schemmer}}]{hinney_unraveling_2021-2}
\bibinfo{author}{\bibfnamefont{J.}~\bibnamefont{Hinney}},
  \bibinfo{author}{\bibfnamefont{A.~S.} \bibnamefont{Prasad}},
  \bibinfo{author}{\bibfnamefont{S.}~\bibnamefont{Mahmoodian}},
  \bibinfo{author}{\bibfnamefont{K.}~\bibnamefont{Hammerer}},
  \bibinfo{author}{\bibfnamefont{A.}~\bibnamefont{Rauschenbeutel}},
  \bibinfo{author}{\bibfnamefont{P.}~\bibnamefont{Schneeweiss}},
  \bibinfo{author}{\bibfnamefont{J.}~\bibnamefont{Volz}}, \bibnamefont{and}
  \bibinfo{author}{\bibfnamefont{M.}~\bibnamefont{Schemmer}},
  \bibinfo{journal}{Physical Review Letters} \textbf{\bibinfo{volume}{127}},
  \bibinfo{pages}{123602} (\bibinfo{year}{2021}).

\bibitem[{\citenamefont{Shen and Fan}(2007)}]{shen_strongly_2007-1}
\bibinfo{author}{\bibfnamefont{J.-T.} \bibnamefont{Shen}} \bibnamefont{and}
  \bibinfo{author}{\bibfnamefont{S.}~\bibnamefont{Fan}},
  \bibinfo{journal}{Physical Review A} \textbf{\bibinfo{volume}{76}},
  \bibinfo{pages}{062709} (\bibinfo{year}{2007}).

\bibitem[{\citenamefont{Le~Kien et~al.}(2004)\citenamefont{Le~Kien, Balykin,
  and Hakuta}}]{le_kien_atom_2004-2}
\bibinfo{author}{\bibfnamefont{F.}~\bibnamefont{Le~Kien}},
  \bibinfo{author}{\bibfnamefont{V.~I.} \bibnamefont{Balykin}},
  \bibnamefont{and} \bibinfo{author}{\bibfnamefont{K.}~\bibnamefont{Hakuta}},
  \bibinfo{journal}{Physical Review A} \textbf{\bibinfo{volume}{70}},
  \bibinfo{pages}{063403} (\bibinfo{year}{2004}).

\bibitem[{\citenamefont{Goban et~al.}(2012)\citenamefont{Goban, Choi, Alton,
  Ding, Lacro{\^u}te, Pototschnig, Thiele, Stern, and
  Kimble}}]{goban_demonstration_2012}
\bibinfo{author}{\bibfnamefont{A.}~\bibnamefont{Goban}},
  \bibinfo{author}{\bibfnamefont{K.~S.} \bibnamefont{Choi}},
  \bibinfo{author}{\bibfnamefont{D.~J.} \bibnamefont{Alton}},
  \bibinfo{author}{\bibfnamefont{D.}~\bibnamefont{Ding}},
  \bibinfo{author}{\bibfnamefont{C.}~\bibnamefont{Lacro{\^u}te}},
  \bibinfo{author}{\bibfnamefont{M.}~\bibnamefont{Pototschnig}},
  \bibinfo{author}{\bibfnamefont{T.}~\bibnamefont{Thiele}},
  \bibinfo{author}{\bibfnamefont{N.~P.} \bibnamefont{Stern}}, \bibnamefont{and}
  \bibinfo{author}{\bibfnamefont{H.~J.} \bibnamefont{Kimble}},
  \bibinfo{journal}{Physical Review Letters} \textbf{\bibinfo{volume}{109}},
  \bibinfo{pages}{033603} (\bibinfo{year}{2012}).

\bibitem[{\citenamefont{Schlosser et~al.}(2002)\citenamefont{Schlosser,
  Reymond, and Grangier}}]{schlosser_collisional_2002-1}
\bibinfo{author}{\bibfnamefont{N.}~\bibnamefont{Schlosser}},
  \bibinfo{author}{\bibfnamefont{G.}~\bibnamefont{Reymond}}, \bibnamefont{and}
  \bibinfo{author}{\bibfnamefont{P.}~\bibnamefont{Grangier}},
  \bibinfo{journal}{Physical Review Letters} \textbf{\bibinfo{volume}{89}},
  \bibinfo{pages}{023005} (\bibinfo{year}{2002}).

\bibitem[{\citenamefont{Vetsch et~al.}(2010)\citenamefont{Vetsch, Reitz,
  Sagu{\'e}, Schmidt, Dawkins, and Rauschenbeutel}}]{vetsch_optical_2010-6}
\bibinfo{author}{\bibfnamefont{E.}~\bibnamefont{Vetsch}},
  \bibinfo{author}{\bibfnamefont{D.}~\bibnamefont{Reitz}},
  \bibinfo{author}{\bibfnamefont{G.}~\bibnamefont{Sagu{\'e}}},
  \bibinfo{author}{\bibfnamefont{R.}~\bibnamefont{Schmidt}},
  \bibinfo{author}{\bibfnamefont{S.~T.} \bibnamefont{Dawkins}},
  \bibnamefont{and}
  \bibinfo{author}{\bibfnamefont{A.}~\bibnamefont{Rauschenbeutel}},
  \bibinfo{journal}{Physical Review Letters} \textbf{\bibinfo{volume}{104}},
  \bibinfo{pages}{203603} (\bibinfo{year}{2010}).

\bibitem[{\citenamefont{Corzo et~al.}(2016)\citenamefont{Corzo, Gouraud,
  Chandra, Goban, Sheremet, Kupriyanov, and Laurat}}]{corzo_large_2016-2}
\bibinfo{author}{\bibfnamefont{N.~V.} \bibnamefont{Corzo}},
  \bibinfo{author}{\bibfnamefont{B.}~\bibnamefont{Gouraud}},
  \bibinfo{author}{\bibfnamefont{A.}~\bibnamefont{Chandra}},
  \bibinfo{author}{\bibfnamefont{A.}~\bibnamefont{Goban}},
  \bibinfo{author}{\bibfnamefont{A.~S.} \bibnamefont{Sheremet}},
  \bibinfo{author}{\bibfnamefont{D.~V.} \bibnamefont{Kupriyanov}},
  \bibnamefont{and} \bibinfo{author}{\bibfnamefont{J.}~\bibnamefont{Laurat}},
  \bibinfo{journal}{Physical Review Letters} \textbf{\bibinfo{volume}{117}},
  \bibinfo{pages}{133603} (\bibinfo{year}{2016}).

\bibitem[{\citenamefont{Rafac et~al.}(1999)\citenamefont{Rafac, Tanner,
  Livingston, and Berry}}]{rafac_fast-beam_1999}
\bibinfo{author}{\bibfnamefont{R.~J.} \bibnamefont{Rafac}},
  \bibinfo{author}{\bibfnamefont{C.~E.} \bibnamefont{Tanner}},
  \bibinfo{author}{\bibfnamefont{A.~E.} \bibnamefont{Livingston}},
  \bibnamefont{and} \bibinfo{author}{\bibfnamefont{H.~G.} \bibnamefont{Berry}},
  \bibinfo{journal}{Physical Review A} \textbf{\bibinfo{volume}{60}},
  \bibinfo{pages}{3648} (\bibinfo{year}{1999}).

\bibitem[{\citenamefont{Mitsch et~al.}(2014)\citenamefont{Mitsch, Sayrin,
  Albrecht, Schneeweiss, and Rauschenbeutel}}]{mitsch_quantum_2014-2}
\bibinfo{author}{\bibfnamefont{R.}~\bibnamefont{Mitsch}},
  \bibinfo{author}{\bibfnamefont{C.}~\bibnamefont{Sayrin}},
  \bibinfo{author}{\bibfnamefont{B.}~\bibnamefont{Albrecht}},
  \bibinfo{author}{\bibfnamefont{P.}~\bibnamefont{Schneeweiss}},
  \bibnamefont{and}
  \bibinfo{author}{\bibfnamefont{A.}~\bibnamefont{Rauschenbeutel}},
  \bibinfo{journal}{Nature Communications} \textbf{\bibinfo{volume}{5}},
  \bibinfo{pages}{1} (\bibinfo{year}{2014}).

\bibitem[{\citenamefont{Abitan et~al.}(2008)\citenamefont{Abitan, Bohr, and
  Buchhave}}]{abitan_correction_2008}
\bibinfo{author}{\bibfnamefont{H.}~\bibnamefont{Abitan}},
  \bibinfo{author}{\bibfnamefont{H.}~\bibnamefont{Bohr}}, \bibnamefont{and}
  \bibinfo{author}{\bibfnamefont{P.}~\bibnamefont{Buchhave}},
  \bibinfo{journal}{Applied Optics} \textbf{\bibinfo{volume}{47}},
  \bibinfo{pages}{5354} (\bibinfo{year}{2008}).

\bibitem[{\citenamefont{Kusmierek et~al.}(2022)\citenamefont{Kusmierek,
  Mahmoodian, Cordier, Hinney, Rauschenbeutel, Schemmer, Schneeweiss, Volz, and
  Hammerer}}]{kusmierek_higher-order_2022-1}
\bibinfo{author}{\bibfnamefont{K.}~\bibnamefont{Kusmierek}},
  \bibinfo{author}{\bibfnamefont{S.}~\bibnamefont{Mahmoodian}},
  \bibinfo{author}{\bibfnamefont{M.}~\bibnamefont{Cordier}},
  \bibinfo{author}{\bibfnamefont{J.}~\bibnamefont{Hinney}},
  \bibinfo{author}{\bibfnamefont{A.}~\bibnamefont{Rauschenbeutel}},
  \bibinfo{author}{\bibfnamefont{M.}~\bibnamefont{Schemmer}},
  \bibinfo{author}{\bibfnamefont{P.}~\bibnamefont{Schneeweiss}},
  \bibinfo{author}{\bibfnamefont{J.}~\bibnamefont{Volz}}, \bibnamefont{and}
  \bibinfo{author}{\bibfnamefont{K.}~\bibnamefont{Hammerer}},
  \emph{\bibinfo{title}{Higher-order mean-field theory of chiral waveguide
  {{QED}}}} (\bibinfo{year}{2022}), \eprint{arXiv:2207.10439}.

\end{thebibliography}

\pagebreak
\newpage
\widetext
\begin{center}
\vspace{1cm}
\textbf{\large Supplemental Material}
\end{center}
\setcounter{equation}{0}
\setcounter{figure}{0}
\setcounter{table}{0}

\makeatletter
\renewcommand{\theequation}{S\arabic{equation}}
\renewcommand{\thefigure}{S\arabic{figure}}
\renewcommand{\thesection}{S\arabic{section}}


\section{Experiment}

\subsection{The experimental setup}

We use an ensemble of laser-cooled cesium (Cs) atoms which are trapped in the vicinity of a \unit[400]{nm} diameter optical nanofiber. The trapping sites surrounding the nanofiber are realized by a two-color optical dipole trap at magic wavelengths for the Cs D2-line~\cite{le_kien_atom_2004-2,goban_demonstration_2012}. We send a blue detuned laser light in a running wave configuration ($P_\mathrm{blue} = \unit[16]{mW}$) at $\lambda= \unit[685]{nm}$ , and a red detuned laser light in a standing wave configuration at $\lambda= \unit[935]{nm}$ ($P_\mathrm{red} = \unit[0.2]{mW}$ per arm). The resulting trapping potential from the two quasilinearly polarized light fields~\cite{le_kien_atom_2004-2} create traps that are deep inside the collisions blockade regime thus limiting the atom number per lattice site to one~\cite{schlosser_collisional_2002-1}. The trapped atoms form two 1D arrays located at a distance of \unit[$\simeq 250$]{nm} from the fiber surface~\cite{vetsch_optical_2010-6,corzo_large_2016-2} and couple to the fiber guided probe field with a coupling constant $\beta = \Gamma_{\mathrm{wg}}/{\Gamma} = 0.007\pm 0.002$. Here $\Gamma$ is the total scattering rate $\Gammatot = \unit[2\pi\times 5.22]{MHz}$ of the Cs D2 line~\cite{rafac_fast-beam_1999} and $\Gamma_{\mathrm{wg}}$ the scattering rate into the waveguide mode. The trapping site distance is incommensurate with $\lambda_{L}=2\pi c/\omega_L$, thus, we expect no coherent effects in back-scattering and only the forward scattered light should be collectively enhanced. The probe light is guided in the nanofiber and couples evanescently to the atoms. It originates from a narrow-band titanium-sapphire laser and is near-resonant with the D2-line transition at $\lambda_L = \unit[852]{nm}$.

\subsection{Filtering the probe light}
In order to separate the weak probe from the strong trapping fields we use a filtering stage that consists of both spectral and polarization filters. In total, we use a polarizing beam splitter, a Dichroic mirror, a volume Bragg grating centered on the D2 line of Cesium, and a $\unit[60]{MHz}$ bandwidth Fabry Perot cavity centered on the probing laser frequency. This allows to separate the signal from the trapping beams (\unit[685]{nm} and \unit[935]{nm}) and reduces the background light originating from spontaneous Raman scattering in the glass fiber material generated by the strong trapping beams. 

\subsection{Sequence}
In each experimental cycle, we first create a cold cloud of Cs atoms in a magneto-optical trap, followed by molasses cooling. The atoms are then loaded into the nanofiber optical lattice.
We probe the atoms on the $6S_{1/2}, F=4 \rightarrow 6P_{3/2}, F'=5$ transition (D2-line). Initially, the atoms are in a combination of all magnetic quantum $m_F$ states. We optically pump the atoms into the $m_F=4$ state by probing during a few microseconds at zero magnetic field~\cite{mitsch_quantum_2014-2}. After the optical pumping, the system behaves as an effective two-level system on the $m_F=4 \rightarrow m_{F'} = 5$ transition. 
Each experimental cycle consists of 120 repetitions of interleaved probing of time $T_{\mathrm{probing}} = \unit[216]{\mu s}$ and molasses cooling for $T_\mathrm{cooling}= \unit[230]{\mu s}$. During each experimental cycle, the atom number slowly decreases over time due to atom losses. This allows us to measure at different atom numbers $N$. The atom number $N$ is determined by measuring the transmission $T = P_{\mathrm{out}}/P_{\mathrm{in}}$ with a sliding average calculated over fifteen probe pulses. 
Simultaneously, for each probing pulse, we record histograms of the photon arrival times with an HBT setup to reconstruct the second-order correlation function $g^{(2)}(\tau)$.   

\subsection{Determination of $\beta$}\label{beta_saturation_meas}
\begin{figure}[h!]
    \centering
    \includegraphics[width=0.5\linewidth]{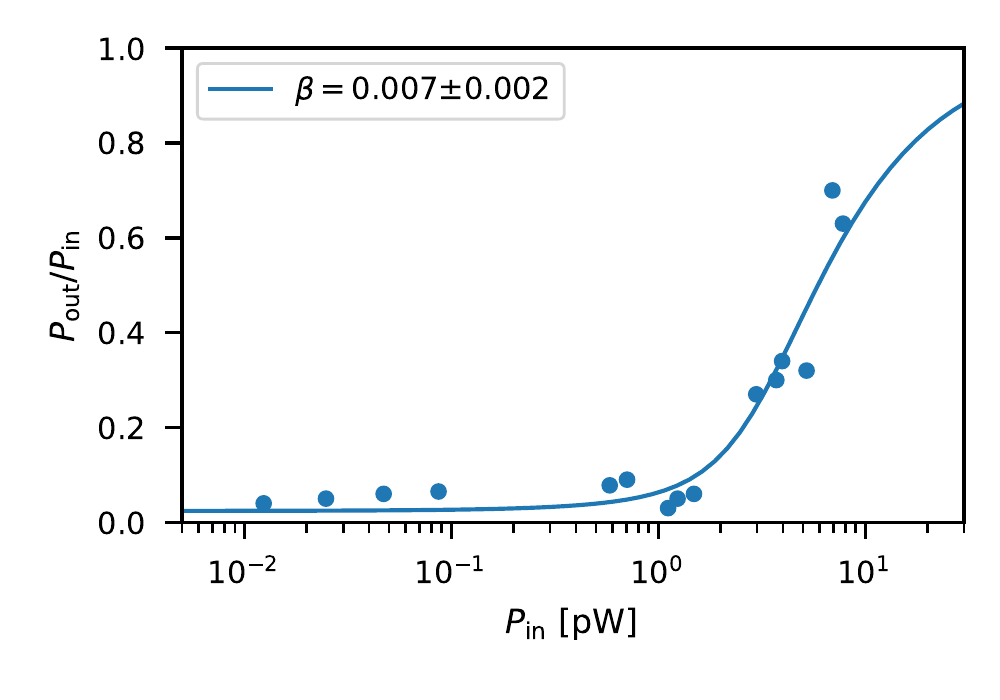}
    \caption{The transmission $T = P_{\mathrm{out}}/P_{\mathrm{in}}$ of the ensemble as function of the input power $P_{\mathrm{in}}$. The solid line is a fit of the data (shown in dots) with free fitting parameters $\beta$ and $N$.}
    \label{fig_sat_measurement_beta}
\end{figure}

We determine the coupling strength $\beta$ in a separate saturation measurement on resonance~\cite{vetsch_optical_2010-6}. We vary the input power between $P_{\mathrm{in}} = \unit[1.7]{pW}$ and $P_{\mathrm{in}} = \unit[1.06]{nW}$ and measure the transmission. The transmission $T$ is then given by an extended Beer-Lambert law~\cite{abitan_correction_2008}:
\begin{equation}
    T = \frac{\mathfrak{W}\left( e^{-4 \beta N + \frac{P_{\mathrm{in}}}{P_{\mathrm{sat}}}} \frac{ P_{\mathrm{in}} }{ P_{\mathrm{sat}} }  \right) }{\frac{P_{\mathrm{in}}}{P_{\mathrm{sat}}}},
\end{equation}
where $\mathfrak{W}$ is the Lambert W function and $s=P_\mathrm{in}/P_\mathrm{sat}$ with $P_\mathrm{sat} = \frac{\hbar\omega_L\Gammatot}{8\beta}$. The fitted data is shown in Fig.~\ref{fig_sat_measurement_beta} from which we obtain $\beta = 0.007\pm0.002$ and $P_\mathrm{sat} = \unit[140\pm40]{pW}$.

\subsection{Control of the amplitude}
\begin{figure}[h!]
    \centering
    \includegraphics[width=0.5\linewidth]{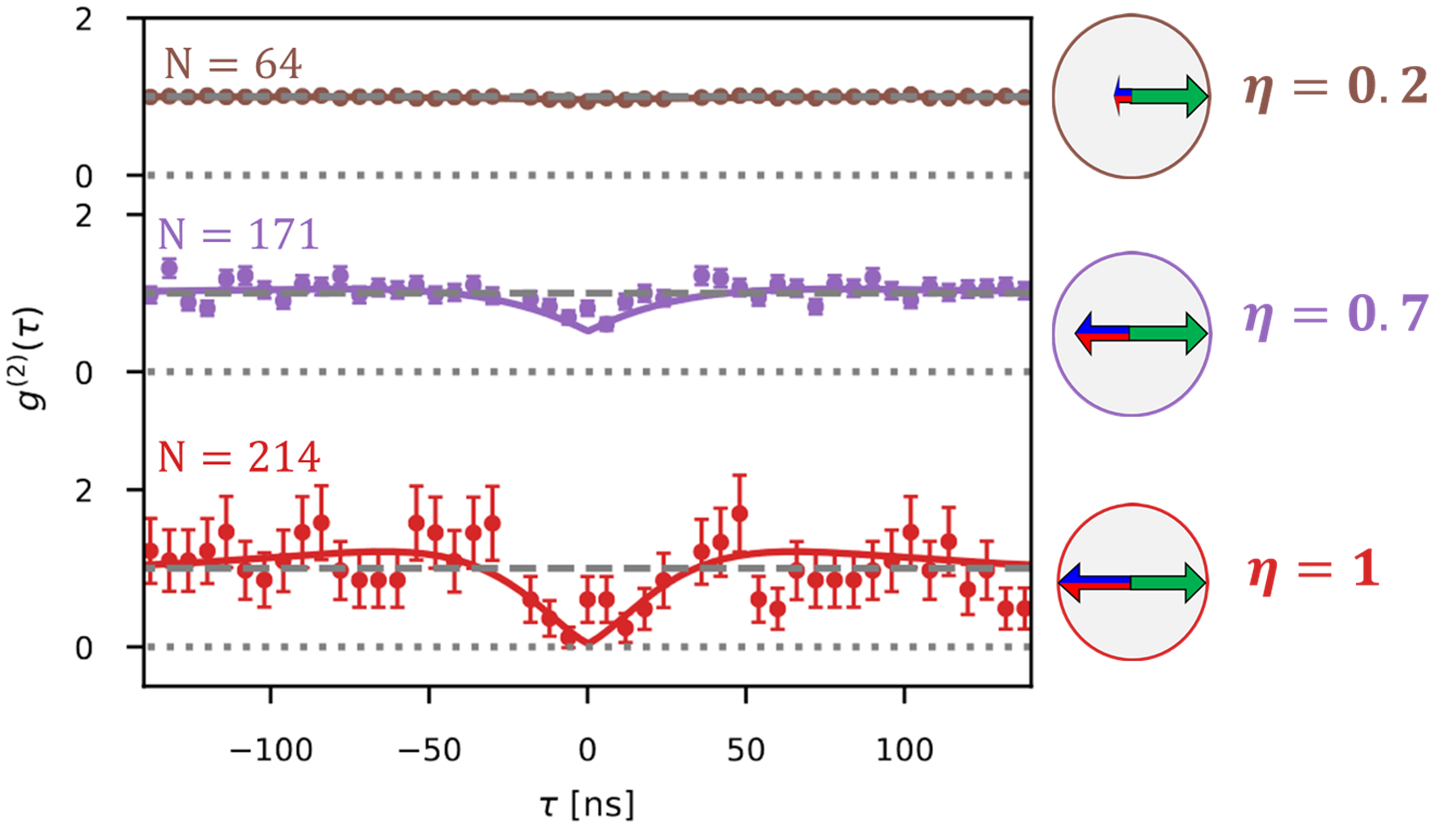}
    \caption{Second-order correlation function $g^{(2)}(\tau)$ for different atom number on a resonant excitation ($\Delta = 0$). From top to bottom the atom number is increasing which translates to an increasing relative amplitude $\eta=\left|\phi(\tau=0)\right|/\left|\alpha^2/2\right|$. The solid line is the model prediction. 
    }
    \label{fig_AmControl}
\end{figure}

The relative amplitude of the coherent and incoherent two-photon component at zero delay $\eta = |\phi(\tau=0)|/|\alpha^2/2|$ can be controlled by tuning the atom number $N$~\cite{prasad_correlating_2020-1}. Figure~\ref{fig_AmControl} shows second-order correlation functions $g^{(2)}(\tau)$ for excitation on resonance ($\Delta = 0$) for different atom numbers. On resonance, the relative phase is $\Delta\varphi(0) = \pi$ resulting in destructive interference between coherent and incoherent photon pairs at $\tau=0$ for any $N$. When gradually increasing the number of atoms, the relative weight of the incoherent photon pairs in the two-photon wavefunction continuously grows. For small $N$ (brown), the transmitted light exhibits a Poissonian photon statistics as the dynamic is governed by the transmitted probe light. For larger $N$ (purple), the presence of incoherent photon pairs becomes visible, as shown by the small antibunching dip. Finally, when reaching the point of balanced coherent and incoherent two-photon amplitudes (red), a strong antibunching is detected. Eventually, for even larger $N$, the two-photon wavefunction is dominated by $\phi$ ($\eta\gg 1$) which leads to a bunching. 

\newpage
\section{Theoretical Model}\label{sec_theory_model}
The theoretical modeling is an extension and simplification of the work in Ref.~\cite{mahmoodian_strongly_2018} to non-resonant excitation. Note that in contrast to~\cite{mahmoodian_strongly_2018,hinney_unraveling_2021-2,kusmierek_higher-order_2022-1} we use a normalization of the coherent two-photon part which differs by a factor $1/2$.
Furthermore, in order to simplify our notation, we normalize the two-photon amplitude $\Psi_2(\tau)$ to the input laser amplitude $\alpha_\mathrm{in}$. Then, the coherent two-photon amplitude follows Beer-Lambert's law with $\alpha^2/2 = t(\Delta)^{2N}$ where $t(\Delta) = 1 - 2\beta/(1 - 2i\Delta/\Gammatot)$ is the transmission coefficient of a single atom.

Calculating the incoherent amplitude $\phi(\tau)$ is more involved since it depends on the collective non-linear scattering of the ensemble. It was first solved for chirally coupled emitters on resonance in~\cite{mahmoodian_strongly_2018}. When considering weakly coupled emitters ($\beta \ll 1$), it can be well described by a simple analytic expression:
\begin{equation}
\phi(\omega) = \sum_{n= 1}^N t({\Delta})^{2(n-1)} \phi_\mathrm{atom}(\omega)\left[ t({\Delta + \omega})t({\Delta - \omega}) \right]^{N-n}
\label{eq_phiN_supp}
\end{equation}
where $\phi_\mathrm{atom}(\omega)$ is normalized amplitude for a given emitter to scatters a correlated photon pair and we introduce the Fourier transform $\phi(\omega) = \int_{-\infty}^{\infty}\mathrm{d}\tau \phi(\tau) e^{i\omega \tau} $. For a single atom, the non-linear two-photon scattering wavefunction is given by~\cite{mahmoodian_strongly_2018}:
\begin{equation}
\phi_\mathrm{atom}(\tau) = - r(\Delta)^2 \, e^{-\Gammatot |\tau|/2} e^{i\Delta |\tau|}
\end{equation} 
that depends quadratically on the linear scattering probability $r(\Delta) = t(\Delta) - 1$,  the term $e^{-\Gammatot |\tau|/2}$ reflects the decrease of the interaction probability with increasing photon distance and the phase factor $e^{i\Delta |\tau|}$. 

Expression~\ref{eq_phiN_supp} can be understood as follows: In forward scattering, $\phi(\omega)$ results from the sum of the individual amplitudes for generating a nonlinear photon pair at each emitter. Moreover, at the $n$-th emitter chain, the coherent two-photon amplitude driving the nonlinear process has been attenuated by the factor $t(\Delta)^{2(n-1)}$  (for $\beta \ll 1$, one can neglect the driving due to the generated nonlinearly scattered photon-pairs), and the $n$-th emitter scatters a correlated photon pair with the amplitude $t(\Delta)^{2(n-1)} \phi_\mathrm{atom}(\omega)$. The photon pairs created through such a nonlinear process exhibit red- and blue-detuned frequency components with respect to the probe frequency. These red- and blue-detuned photons generated from the non-linear scattering are then transmitted with the transmission coefficient $\left[ t({\Delta + \omega})t({\Delta - \omega}) \right]^{N-n}$ by the remaining $N-n$ atoms in the chain.


From $\phi(\omega)$ and the attenuated probe field $\alpha^2/2$, we can then calculate the normalized second-order correlation function $g^{(2)}(\tau)$~\cite{mahmoodian_strongly_2018}
\begin{equation}
    g^{(2)}(\tau)  = \frac{|\alpha^{2}/2 + \phi(\tau)|^2}{|\alpha|^{4}} + \mathcal{O}(s) \label{eq_g2_supp}.
\end{equation}
Eq.~\eqref{eq_g2_supp} is valid within the low saturation regime $s\ll 1$ where $s=P_\mathrm{in}/P_\mathrm{sat}$ is the saturation parameter of the first atom. This condition is experimentally well fulfilled for our experimental probe power $P_\mathrm{in} = \unit[2]{pW}$ ($s=0.03$). 

\newpage
\section{Numerical analysis}
\subsection{Comparison Data vs Model prediction}
\begin{figure}[h!]
    \centering
    \includegraphics[width=0.8\linewidth]{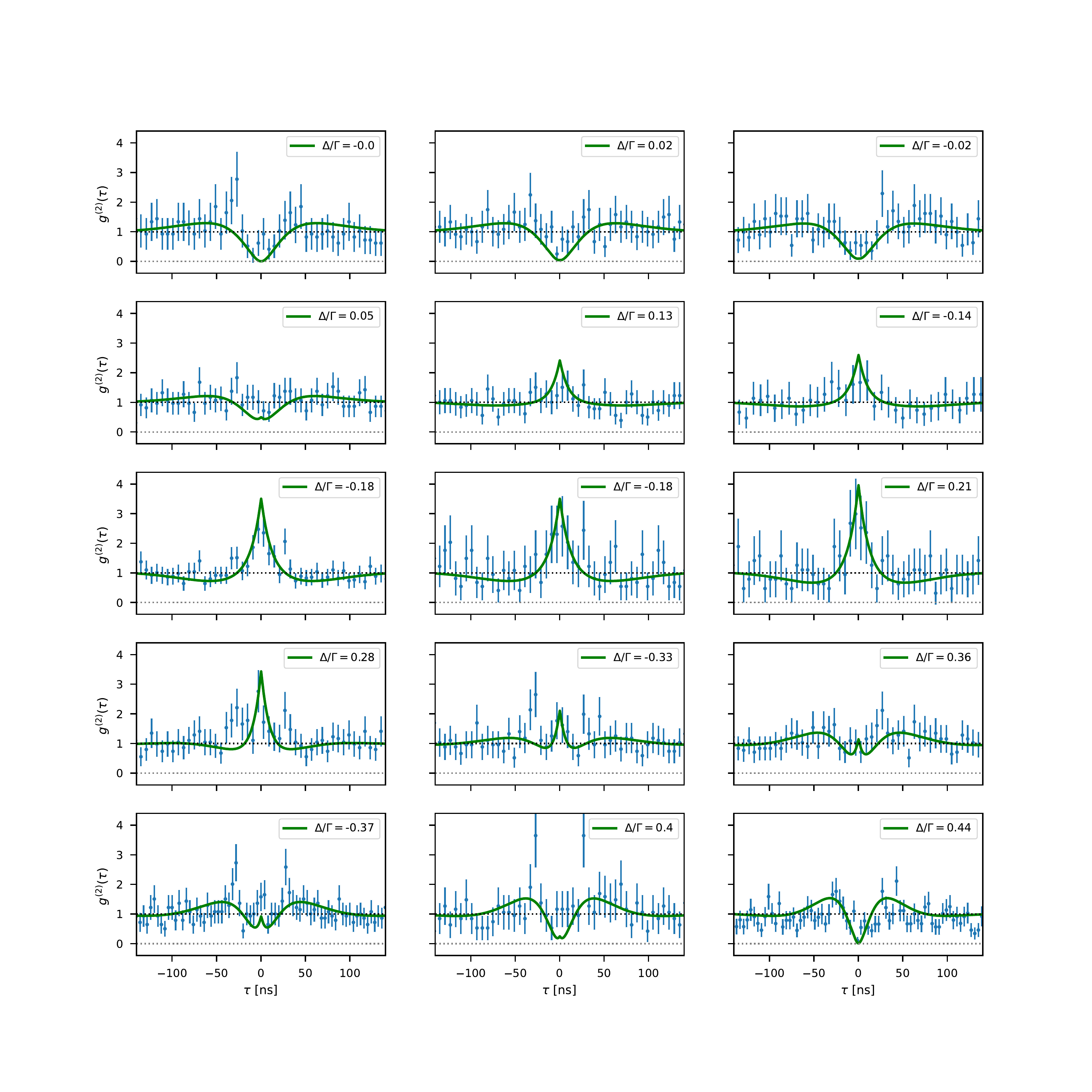}
    \caption{Measured (blue data) and model prediction (green line) second-order correlation functions $g^{(2)}(\tau)$ for equal amplitude of coherent and incoherent component at zero delay in ascending order in $\Delta\varphi(0)$. The theoretical predictions (green line) do not take into account experimental imperfections, see section \ref{sectionImoerfections}.  }
    \label{fig_2D_model}
\end{figure}

\newpage
\subsection{Maximum likelihood fit}
\begin{figure}[h!]
    \centering
    \includegraphics[width=0.8\linewidth]{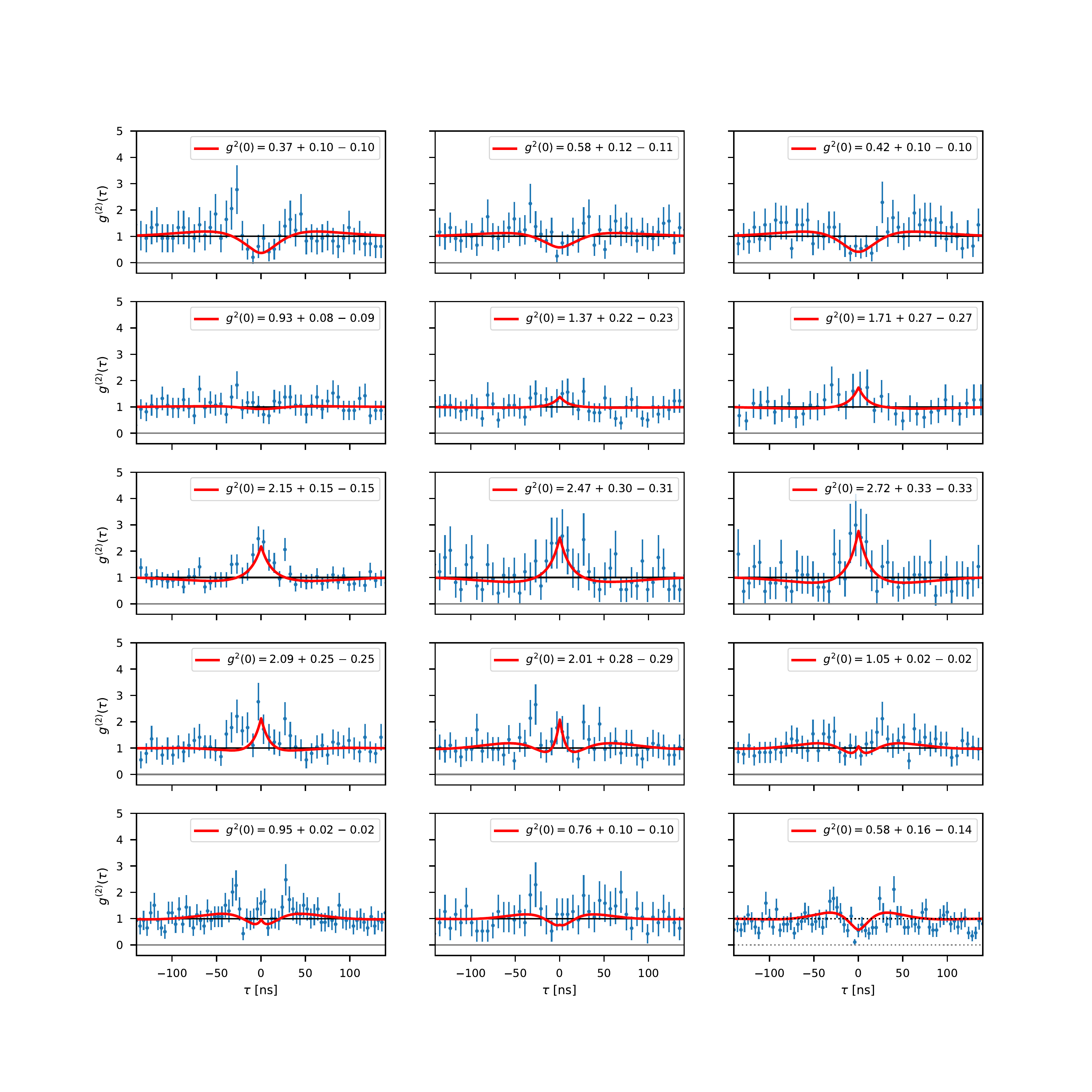}
    \caption{The second-order correlation function $g^{(2)}(\tau)$ on ascending order in $\Delta\varphi(0)$ (the corresponding detuning is indicated in Figure~\ref{fig_2D_model} where the plots are shown in same order). The red solid light is the MLE fit. }
    \label{fig_2D_fit}
\end{figure}

The measured second-order correlation functions are fitted with the theoretical model together with a visibility factor $C$, which takes into account a reduced contrast due to experimental imperfections. Since the count rates are small, we use a maximum likelihood estimation (MLE) method instead of a $\chi^2$-fit. We maximize the likelihood $\mathcal{L}(C)= \prod_i(p\left(G^{(2)}(\tau_i)),G_\mathrm{exp}^{2}(\tau_i) \right)$ where $G_\mathrm{exp}^{2}(\tau_i)$ is the expected coincidence count and $p(x,\lambda)$ the Poissonian distribution. The expected coincidence count is inferred from the previously introduced model for $g_\mathrm{model}^{(2)}(\tau)$ with a reduced visibility such that $g_\mathrm{model}^{(2)}(\tau) =  C \left\lbrace g^{(2)}(\tau)-1 \right\rbrace + 1$. The fit results, together with the corresponding fit curves, are shown in Fig.~\ref{fig_2D_fit}. From the MLE, we infer the positive and negative confidence intervals on $g^{(2)}(0)$, which are shown in the legends. 

\newpage
\section{Effect of parameter uncertainty and data binning }\label{sectionImoerfections} 
In order to adjust our setup to the point where the coherent and incoherent two-photon amplitude exactly matches ($\eta=1$), a precise knowledge of the system's parameters $\Delta$, $N$ \& $\beta$ is necessary. In the following, we summarize the errors in these quantities and how they affect our measured interferences finges. 

\paragraph{Atom number:}

\begin{figure}[h!]
    \centering
    \includegraphics[width=\linewidth]{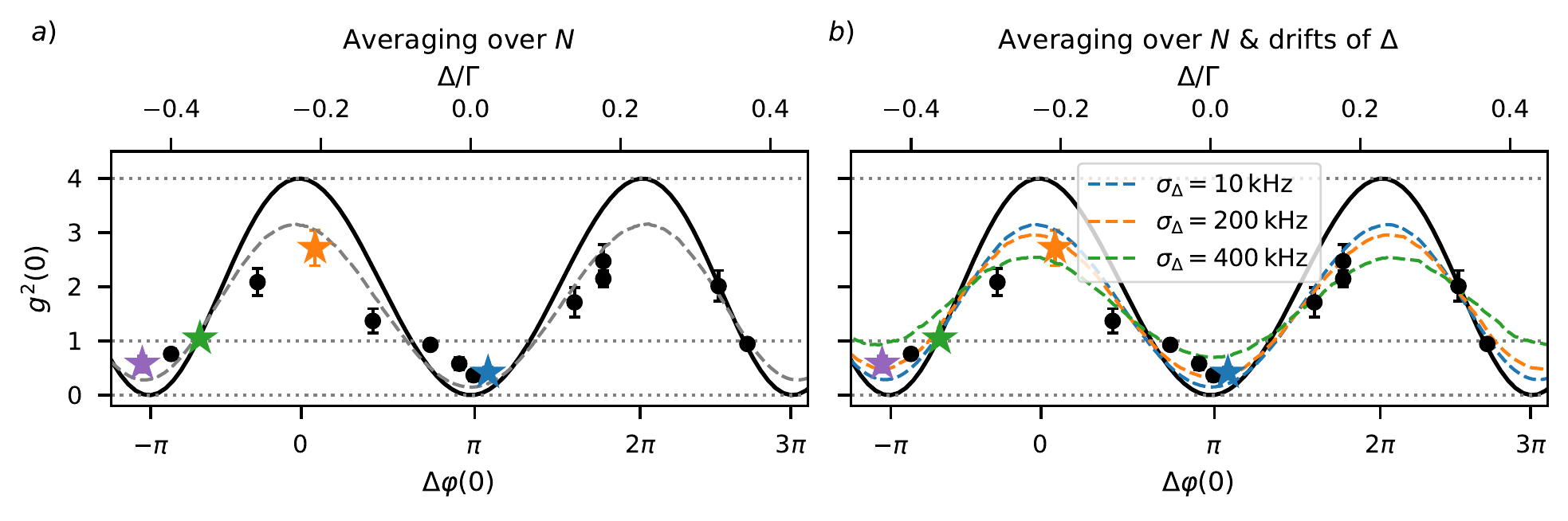}
    \caption{The effect of the experimental imperfections on the two-photon interference visibility. The data points and theory prediction in black are the same as in Fig.~\ref{fig_am}. 
    a) The effect of the averaging over a finite range of $N$ (or OD-binning) is shown by the gray dashed which is close to the observed reduction of the interference fringes.
    b) The combined effect of OD-binning and uncertainty in $\Delta$ for different values of $\sigma_\Delta$. }
    \label{fig_am_N_uncrt}
\end{figure}

We attribute the main contribution for a reduced interferometer visibility (see Fig.~\ref{fig_am}) to the averaging over a range of atoms number $N$. More precisely, in our analysis, we group runs with similar OD and average over a small range of $\Delta OD \simeq 1.7$. Within each OD-bin, the probability distribution $p(\mathrm{OD})$ is approximately flat.
In order to estimate the effect of OD-binning on the visibility, we perform a Monte-Carlo simulation using these parameters. The effect on the interference fringes is shown by the dashed gray line in Fig.~\ref{fig_am_N_uncrt} a). 

\paragraph{Atom-laser detuning:}
We estimate that fluctuations in $\Delta$ are mainly due to drifts in our laser locking system during the $\simeq 8h$ of measurement time for each data point. As a quantitative estimation of these drifts is difficult, we estimated this effect by performing the same Monte-Carlo simulation as for $N$ (Fig.~\ref{fig_am_N_uncrt} a)) and adding an additional Gaussian distribution of detunings of width $\sigma_\Delta $. The resulting average $g^{(2)}(0)$ values are shown in Fig~\ref{fig_am_N_uncrt} b) for different values of $\sigma_\Delta$ (dashed lines). Comparing these curves to our experimental data suggests that fluctuations on the order of $\sigma_\Delta =\unit[200]{kHz}$ are compatible with our experimental observations. Such small drifts are on the order of the laser linewidth and thus are expected over such a long measurement duration.

\paragraph{Coupling constant:}
Additional uncertainty stems from the imprecise knowledge of the coupling strength $\beta = 0.007\pm 0.002$ as described in Sec.~\ref{beta_saturation_meas}. Assuming Gaussian fluctuations of $\beta$ with a standard deviation corresponding to the measurement uncertainty,
we performed the same type of Monte-Carlo simulation, which showed that fluctuations in $\beta$ have a negligible effect on the visibility compared to the previous parameters $N$ and $\Delta$.

\newpage
\section{Two-photon interferometer}
\subsection{Modifying phase and amplitude} 
Our system can be interpreted as an effective interferometer for the two-photon component as sketched in Fig.~\ref{fig_interferometer} b). One arm contains the coherent photon pairs $\alpha^2/2$ and the other arm the incoherent part $\phi$. In principle, both the phase difference $\Delta\varphi(0)$ and the amplitude $\eta$ are a function of $N$
 and $\Delta$. 
For $\Delta/\Gamma \ll 1$, and close to the point where the coherent and incoherent two-photon amplitudes are balanced ( $|\alpha/2| = |\phi(\tau = 0)|$), one can show that $\Delta$ essentially modifies the relative phase $\Delta\varphi(0)$, while $N$, the relative amplitude $\eta$. More precisely, one can show that the following approximate scaling holds:
\begin{align}
\Delta \varphi(0) & = \mathrm{Arg}\left\lbrace\frac{\phi(0)}{\alpha^2/2}\right\rbrace \simeq\pi + 8 \tilde{N}(0) \frac{\Delta}{\Gamma} \label{eq_am_matching_phase_approx}\\
\eta &= \frac{|\phi(0)|}{|\alpha^2/2|} \simeq |t(\Delta)^2|^{\tilde{N}(\Delta)-N}\propto  e^{N}.\label{eq_am_matching_eta_approx}
\end{align}
where $\tilde{N}(\Delta)$ is the atom number for which coherent and incoherent two-photon magnitude are balanced at zero delay  ($|\alpha/2| = |\phi(\tau = 0)|$).
\begin{figure}[h!]
    \centering
    \includegraphics[width=\linewidth]{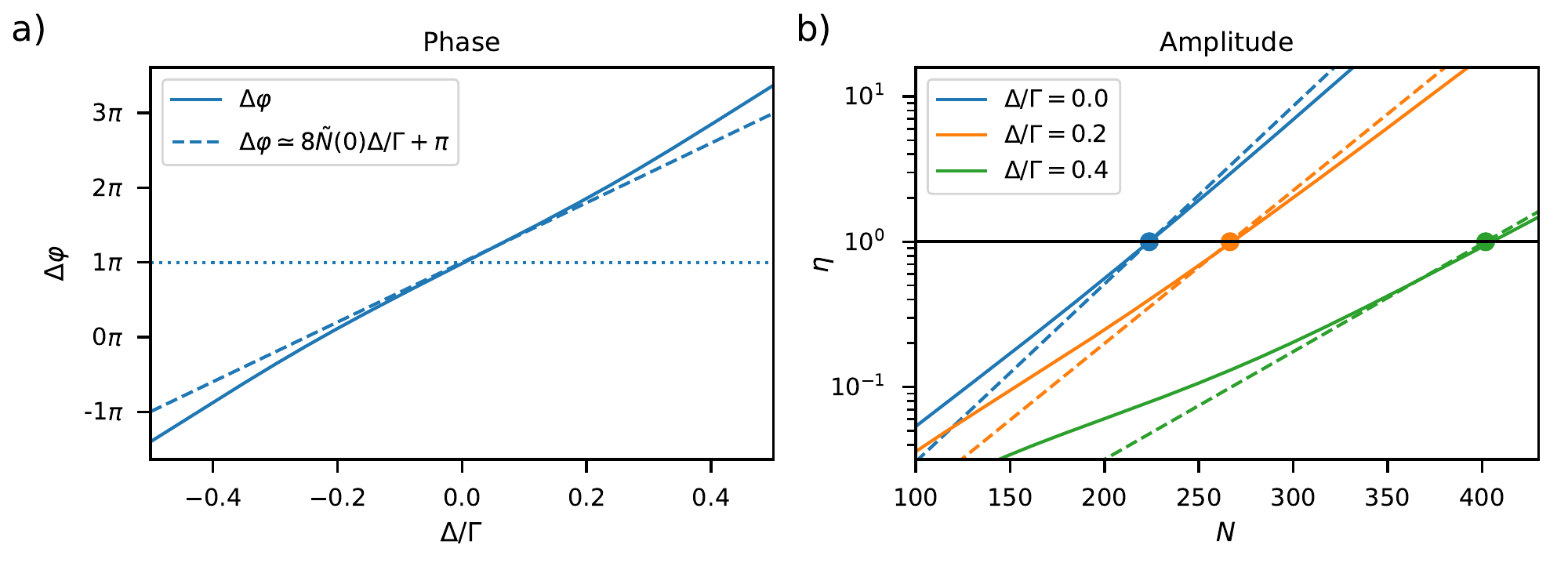}
    \caption{The control parameters of the two-photon interferometer a) The phase $\Delta \varphi(0)$ at amplitude matching ($\eta=1$) as a function of detuning $\Delta$ (solid line). For small detuning $\Delta\ll\Gamma$, $\Delta \varphi(0)$ varies almost linearly with $\Delta$ as predicted by Eq.~\eqref{eq_am_matching_phase_approx} (dashed line). 
    b) The relative amplitude of the two-photon component $\eta = |\phi(\tau=0)|/|\alpha^2/2|$ as a function of $N$ (solid line). Around $\eta=1$  (indicated by the dots), $\log(\eta)$ varies linearly with $N$ as predicted by Eq.~\eqref{eq_am_matching_eta_approx} (dashed lines).
    }
    \label{fig_AM_plot_tildeN_zeta}
\end{figure}
As shown in Fig.~\ref{fig_AM_plot_tildeN_zeta} a) where we compare this approximation to the full calculation based on Sec.~\ref{sec_theory_model}, the phase $\Delta \varphi(0)$ indeed varies almost linearly with $\Delta$ within the experimentally relevant range $|\Delta| \lesssim 0.5 \Gammatot$. Therefore the interference fringes in Fig.~\ref{fig_am} are very close to sinusoidal modulation with respect to $\Delta$. For larger detuning ($|\Delta| \gg 0.5 \Gammatot$), $g^{(2)}(0)$, still oscillates between 0 and 4, but the period of the oscillation decreases as $\Delta$ increases.

At the same time, close to the point of amplitude matching, $\log(\eta)$ varies linearly with $N$ as shown in Fig.~\ref{fig_AM_plot_tildeN_zeta} b). In our two-photon interferometer, it is possible to reach amplitude matching for any detuning $\Delta$ due to the exponential decay of the coherent component $|t(\Delta)^{2N}|$, and the initial growth followed by sub-exponential decay of the incoherent component $|\phi(\tau=0)|$. 

\subsection{Interference fringes at different time-delay $\tau$}
 \begin{figure}[h!]
    \centering
    \includegraphics[width=\linewidth]{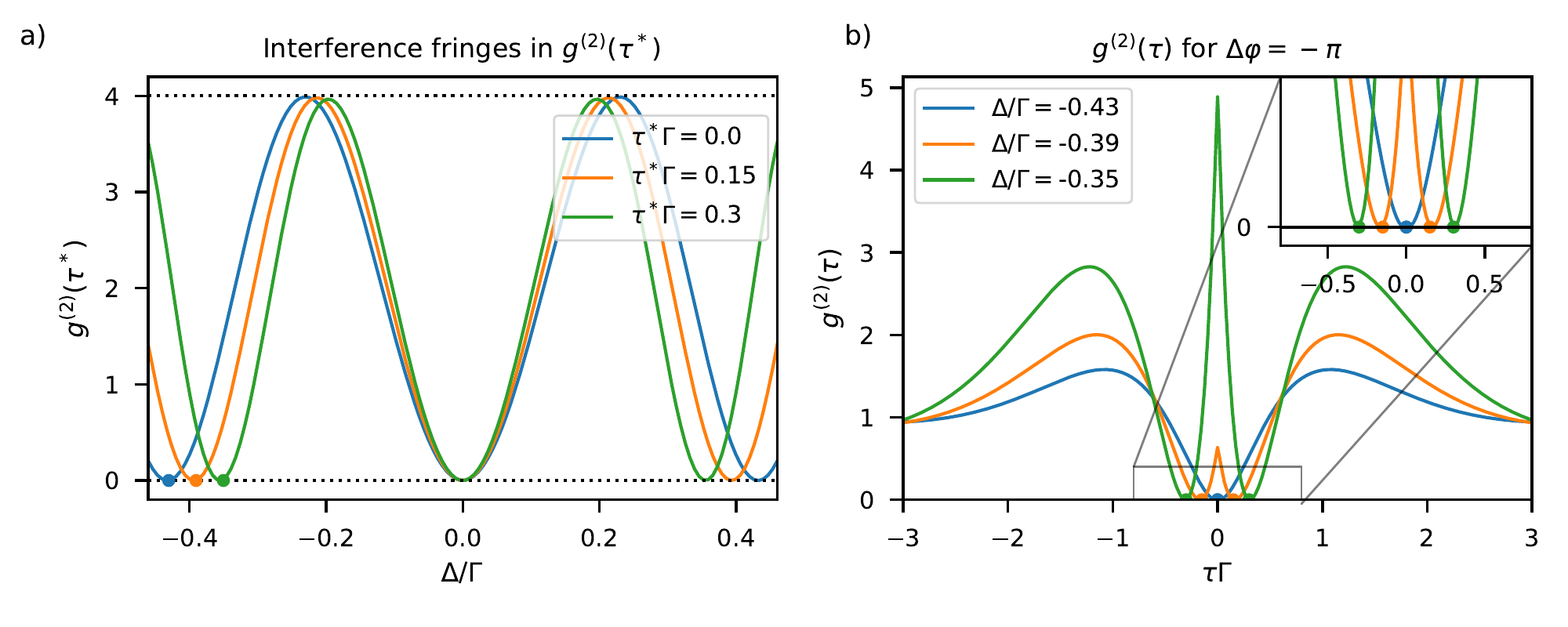}
    \caption{a) Interference fringes for different values of $\tau^*$ as a function of $\Delta$ at amplitude matching $\eta(\tau^*)=1$. b) Three second-order correlation functions $g^{(2)}(\tau)$ that exhibit perfect destructive interference ($\Delta\varphi (\tau^*)=-\pi$) at the three times $\tau^*$ indicated in a). The colors in both panels correspond to the same $\tau^*$. }
    \label{fig_AM_matching_taus}
\end{figure}
While in this article, we study interference in the coincidence rate  ($\tau=0$), the same interference phenomenon can occur for any  $\tau^*$. In principle, the two-photon interferometer allows one to measure interference fringes in  $g^{(2)}(\tau^*)$ when adapting the amplitude condition  $\eta(\tau^*) = |\phi(\tau^*)|/|\alpha^2/2| =1$. When choosing $\tau^*\neq  0$, the number of atoms needed to fulfill $\eta(\tau^*)=1$ increases, and the oscillation frequency of the fringes is slightly faster in terms of $\Delta$ as shown in Fig.~\ref{fig_AM_matching_taus} a). In practice, measuring the interference fringes for different $\tau^*$ is challenging since it requires measuring at higher optical depths and thus lower transmission. Furthermore, faster oscillations occur in $g^{(2)}(\tau^*)$, which are increasingly difficult to resolve temporally as shown in Fig.~\ref{fig_AM_matching_taus} b).

\newpage

\end{document}